\documentclass[%
 aip,cha,
 amsmath,amssymb,superscriptaddress,
reprint,%
]{revtex4-2}

\usepackage{graphicx}
\graphicspath{ {images/}} 
\usepackage{dcolumn}
\usepackage{bm}
\usepackage[utf8]{inputenc}
\usepackage[T1]{fontenc}
\usepackage{mathptmx}
\usepackage[caption=false]{subfig}

\begin{document}

\title{Classical-to-Topological Transmission Line Couplers}

\author{Robert J. Davis}
\email[Corresponding authors: ]{rjdavis@ucsd.edu; dsievenpiper@ucsd.edu}
\affiliation{Electrical Engineering Department, University of California, La Jolla, California 92093, USA}
\author{Dia'aaldin J. Bisharat}
\affiliation{Electrical Engineering Department, University of California, La Jolla, California 92093, USA}
\affiliation{Photonics Initiative, Advanced Science Research Center, City University of New York, New York, New York 10031, USA}
\author{Daniel F. Sievenpiper}
\affiliation{Electrical Engineering Department, University of California, La Jolla, California 92093, USA}

\date{\today}

\begin{abstract}
Recent advances in topologically robust waveguiding for electromagnetic systems have presented opportunities for improving practical photonic and microwave devices. To bring this rich area of physics within the reach of application, it is critical for such systems to be interfaced with classical, continuous waveguiding and transmission line technology. This Letter presents a compact, highly efficient transition from a classical metallic transmission line to a topologically nontrivial line wave emulating the quantum spin Hall effect. A zero-gap antipodal slot line is used as the starting transmission line, which is then coupled to the topological metasurface via a field matching procedure. Additional modifications to the interface between the two structures to eliminate unwanted edge coupling improves transmission further. A simulated loss analysis isolates the effect of the transitions from the rest of the structure, showing a loss contribution of only 2.1\% per classical-to-topological conversion. Using the transition, a quantitative characterization of the robustness of common topologically protected devices is presented. This design lays the foundation to integrate topologically robust metasurface transmission lines to traditional systems, opening the door to future uses of such structures in systems.
\end{abstract}

\maketitle

Photonic topological insulators (PTIs) are an exciting option for advanced manipulation of electromagnetic modes, but there is a present lack of understanding on how such modes interact with traditional (classical) electromagnetic modes. PTIs, generally implemented via metamaterials or metasurfaces, permit the flow of electromagnetic (EM) energy only along a boundary of a nontrivial bulk \cite{ozawa_topological_2019}. A chief benefit of such modes is their inherent robustness to disorder, where backscattering is effectively eliminated for fabrication defects and sharp bends of the propagation channel. Despite these benefits, there are currently limited means of efficiently coupling energy into any of these topologically protected devices. Standard experimental methods generally rely on local point source excitations placed near the boundary \cite{wang_observation_2009}\cite{chen_experimental_2014}, or via direct (non-optimal) coupling to waveguides or other transmission lines \cite{he_silicon--insulator_2019}. This allows for verification of nontriviality, but is incompatible with any practical application and restricts quantitative comparisons to classical devices. This Letter presents a design and general methodology for an optimized abrupt coupler between a classical microwave transmission line and a topological edge mode, with corresponding analyses on loss mechanisms and robustness, which will permit a greater understanding of how these exotic topological modes interact with ordinary EM modes. It is an expansion of the authors' preliminary results of \cite{davis_efficient_2020}.

Previous efforts have been made to couple energy from traditional transmission lines to spoof surface plasmon polariton (SPP) modes \cite{ma_broadband_2014}\cite{kianinejad_design_2015}, as well as to 1-D line waves \cite{xu_adiabatic_2019}, and a recent study \cite{gentili_towards_2019} has demonstrated a unidirectional source for a parallel-plate waveguide PTI with high efficiency. However, there are as yet no methods for efficient coupling between surface-wave engineered PTI modes, which represent a promising platform for application, and planar transmission lines. In both \cite{ma_broadband_2014} and \cite{xu_adiabatic_2019} adiabatic transitions are used to convert continuously homogeneous transmission lines into spatially periodic modes, which allows for flexibility in matching the transverse mode profile and/or the propagation constant to achieve high transmission. Such methods are attractive for their flexibility to the initial choice of source, but are limited by the length of the transition and by unwanted effects of radiation from flaring ground planes. In \cite{gentili_towards_2019}, the source is judiciously chosen to permit a unidirectional excitation, and impedance matching methods are employed to minimize reflections. Such techniques can achieve very high efficiency, but are complicated by the difficulty in choosing an appropriate definition for impedance, and by the number of potential geometrical degrees of freedom used to match \cite{lawrence_modeling_2013}.

\begin{figure}
	\centering
	\includegraphics[width = \linewidth]{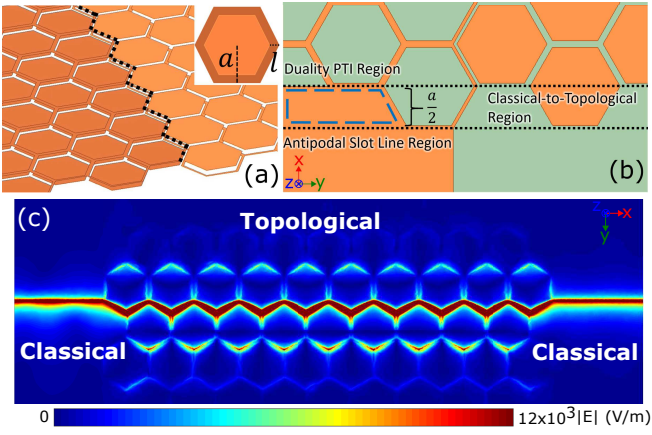}
	\caption{(a) EM duality-based spin PTI structure used, with the dielectric removed. The black dashed line shows the path the 1D line wave follows. Inset is the unit cell used, with the lattice period $a=7$ mm and frame width $l=1$ mm. (b) Overview of the presented classical-to-topological mode coupling region. Orange denotes metal, while the green is the dielectric support (Rogers RT/duroid R 5880 of thickness $t = 0.787$ mm). The blue dashed region shows the modification to the interface, as discussed in the text. (c)  Electric field profile 0.1 mm above the top of the structure (10 unit cells long by 6 wide), with the classical antipodal slot line sections connected at either side. }
	\label{fig1}
\end{figure} 

A particularly attractive option for practical implementation of PTIs is given in \cite{bisharat_electromagnetic-dual_2019}, which employs a stack of two patterned metallic metasurfaces between a dielectric spacer, easily implemented via standard PCB fabrication methods. Such a platform for PTIs combines two key ideas: 1. EM duality \cite{khanikaev_photonic_2013} and 2. the pseudospin degree of freedom for photons. As analyzed in \cite{bisharat_guiding_2017}-\nocite{horsley_one_2014}\cite{kong_analytic_2019}, when a 2D material characterized by a capacitive surface impedance $Z_c$ is placed next to another of complementary inductive surface impedance $Z_i$, there can exist an EM mode at the interface. By necessity, such a mode is tightly confined to the interface and can therefore be considered a 1D mode, or line wave. The condition for complementary surface impedances can be accomplished via EM duality, where $\epsilon=\mu$, which is possible to implement via Babbinet's principle in metallic surfaces. \cite{bisharat_guiding_2017} demonstrates this via a sheet of metal matches (capacitive) interfaced with the dual surface of a connected metal frame (inductive).

Line waves show surprising robustness, but are limited by their lack of a bandgap. The second property, the pseudospin degree of freedom of photons, is where the connection to PTIs comes in. As demonstrated in \cite{bisharat_electromagnetic-dual_2019}, by bianisotropically coupling modes from two such 1D line waves, it is possible to induce a topologically nontrivial mode at the interface, which emulates the quantum spin Hall effect \cite{kane_quantum_2005}. To do so, the lattice is chosen to be hexagonal, such that there is guaranteed degeneracy at the K(K') point of the Brillouin zone. To induce the PTI mode, this degeneracy is broken by placing a second sheet of complementary metasurfaces on top of the first, but flipped with respect to the interface (i.e., patches above frames, frames above patches), as shown in Fig. \ref{fig1}(a). When the two sheets are close enough to hybridize the TE-dominant patch surfaces and TM-dominant frame surfaces, there is a bianisotropic coupling that opens a bandgap near K(K'), where it can be shown that the hybrid electromagnetic modes $\psi^{\uparrow\downarrow} \equiv E\pm H$ exhibit a non-trivial spin Chern number \cite{ma_guiding_2015}. Hence, such a structure supports highly robust (though reciprocal) PTI edge modes. 

\begin{figure*}
	\centering
	\includegraphics[width = \textwidth]{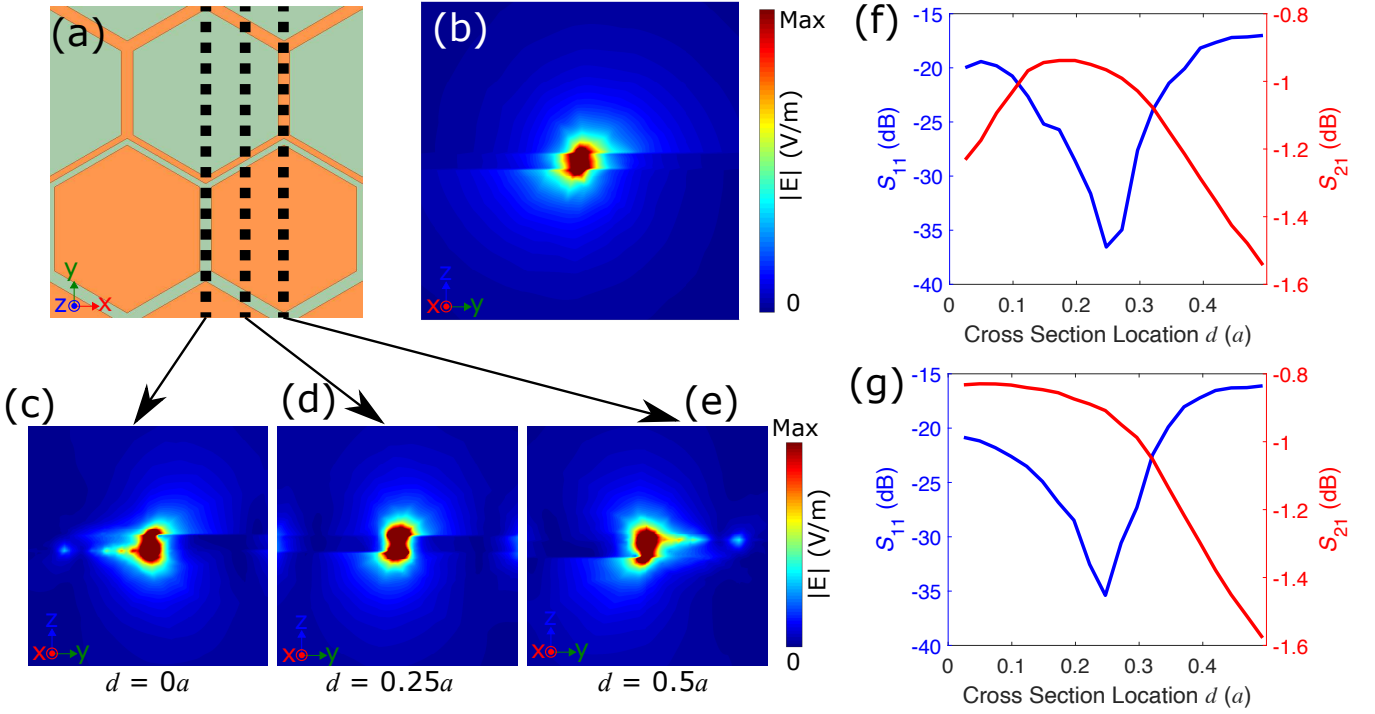}
	\caption{Effects of spatial inhomogeneity coupling. (a) Reference plane for (c)-(e), showing the longitudinal locations chosen. Cross sectional electric field profiles of the (b) ASL (uniform along x) and (c)-(e)  PTI along half a unit cell, demonstrating high variation. Note that at $d=0.25a$ the profile is a near match for the ASL shown in (b). (f) Scattering parameters for the ASL directly interfaced with the PTI vs coupling point on the unit cell $d$. Here, 0 denotes the outer edge of the unit cell. Note that $S_{11}$ is minimized when the field matching condition is closest, at $d=0.25a$, as predicted by the field profiles. (g) S parameters for the modified interface case, as discussed in the text. Here $S_{11}$ remains unchanged, but $S_{21}$ at $d = 0a$ is improved by reducing edge coupling.}
	\label{fig2}
\end{figure*}

The spatial variation in the unit cell of the PTI gives rise to rapidly varying mode profiles, as depicted in Fig. \ref{fig2}. Along the unit cell the electric field changes from highly elongated at the edges to tightly confined vertically between the two inductive layers. This suggests that the location where a traditional transmission line is coupled to the PTI will have a large influence for such structures. A natural choice of traditional transmission line is the antipodal slot line (ASL, the mode profile for which is shown in Fig. \ref{fig2}(b)), commonly used for feeding planar Vivaldi antennas \cite{langley_balanced_1996}. Such a slot line has the E-field concentrated between the edge of the top and bottom metal layers and closely matches the PTI field profile at 1/4 and 3/4 of a unit cell. 

Combining this choice of transmission line with the feed point location as determined by the field profile along the unit cell results in a good match. This is shown in Fig. \ref{fig2}(f) by sweeping the connection point between the ASL and the unit cell edge. For all further analysis the results are given for a 10 unit cell long (i.e., along the propagation direction) and 6 cell wide (3 cells on either side of the interface) sample, with the coupler being equal on both sides. Note that  3 unit cells in the "bulk" is reasonable due to the rapid field decay (see the Supplemental Material). Fig \ref{fig2}(f) shows there is a dip in the simulated reflection ($S_{11}$) at the $d = 0.25a$ connection point, with a corresponding peak in transmission ($S_{21}$). 

Another difficulty is the  mismatch that occurs between the hexagonal lattice of the PTI and straight transmission lines. The regions between the abrupt transition create small cavities along the edge, which can be  undesirably coupled into by the ASL. This is also the cause of the slight shift in transmission maximum for the case shown in Fig. \ref{fig2}(f). To reduce such effects, it is sufficient to metalize a half unit cell past the interface cells, thereby increasing the distance between the first potential edge coupling site and the ASL mode, shown the blue dashed region in Fig. \ref{fig1}(b) (the modified interface case). This causes a slight shift in the optimal location from $0.25a$ to $0a$, as indicated in Fig. \ref{fig2}(g), where the improvement in $S_{21}$ comes as a result of reducing this unwanted boundary coupling, rather than from the field match. Performing the same procedure on more rows can reduce such effects further, but the exponential decay of the classical mode causes these coupling effects to decrease rapidly with distance, so doing so is unnecessary. Fig. \ref{fig1}(c) shows the electric fields 0.1 mm above the top sheet for this case. 

The proposed design uses a balanced transmission line to achieve the field match, and as such is readily converted into many other transmission lines. In such cases, an adiabatic transition is frequently used \cite{pozar_microwave_2011}. Alternatives can include stub matches or other impedance transformations. For the purposes of generality the performance here was characterized in terms of the ASL itself, neglecting contributions from other transmission line conversions. For the experimental data presented, an exponential taper from a microstrip line to the antipodal mode was designed to allow for straightforward connection between the vector network analyzer (VNA, which uses standard SMA connectors) and the device. The taper was partially optimized via a genetic algorithm, but otherwise was not considered further. 

\begin{figure}
	\centering
	\includegraphics[width = \linewidth]{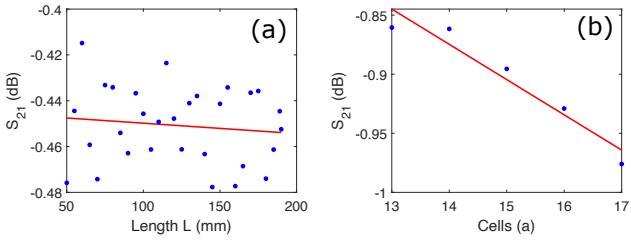}
	\caption{(a) Estimated radiation losses of the PTI section $L_{rad,PTI}$, extracted by sweeping the number of unit cells and calculating the slope, assuming a linear relationship. (b) Estimated radiation losses of the ASL, $L_{rad,anti}$, via simulating a length sweep.}
	\label{fig3}
\end{figure}

To determine the efficiency of the proposed transition, the losses can be broken into several parts and analyzed separately via simulation. Namely, the magnitude of the loss is the sum of the dielectric losses, the radiation losses (both in the PTI and the antipodal sections), and the two transitions,
\begin{equation}
L = L_{d} + L_{rad,PTI} + L_{rad,anti} + L_{tran}.
\end{equation}
Dielectric losses $L_{d}$ can be extracted by setting the loss tangent of the material to zero and subtracting the results. To determine the radiation losses of the PTI section the number of cells is swept while keeping the dielectric lossless. Likewise, the radiation loss from the antipodal sections can be found by sweeping their length. After these are found, adding them to the total loss will give the losses incurred by the transitions themselves. Results of this analysis are given in Table \ref{tab 1}, with details of each component shown in Figs. \ref{fig3}(a)-(b).

The current design exhibits a maximum $S_{21}$ of -0.833 dB for the entire structure operating at 16.2 GHz, with a corresponding $S_{11}$ of -21.0 dB. From the loss analysis, the contribution from the classical-to-topological mode conversion $L_{tran}$ is only 0.181 dB (2.1\%) per transition.
\begin{table}[h]
	\begin{center}
		\caption{Loss mechanisms and magnitudes} \label{tab 1}
		\begin{tabular}{|c|c|}
			\hline 
			Mechanism & Magnitude \\ 
			\hline 
			Dielectric ($L_{d}$) & 0.174 dB \\ 
			\hline 
			Radiative, PTI ($L_{rad,PTI}$) & 0.0298 dB/cell \\ 
			\hline 
			Radiative, Antipodal ($L_{rad,anti}$) & 0.000454 dB/cm \\ 
			\hline 
			Transitions x2 ($L_{tran}$) & 0.181 dB \\ 
			\hline 
		\end{tabular} 
	\end{center}
\end{table}

\begin{figure}
	\centering
	\includegraphics[width = \linewidth]{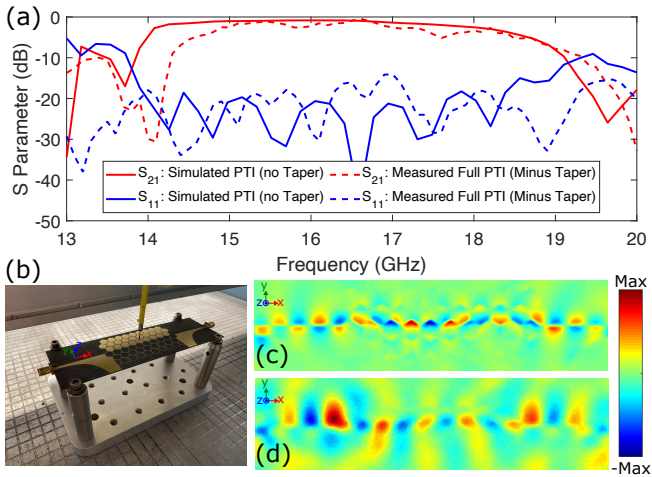}
	\caption{(a) Classical-to-topological conversion performance vs frequency. For the experimental result, an exponential taper was used to connect to the SMA connectors of the VNA (shown in (b)), but to isolate the actual mode conversion performance, the simulation was performed without the taper, and the experimentally measured response of the taper subtracted from that of the full structure. Both simulated and measured data include all forms of losses otherwise. The measured transmission is better than -3 dB within 15-18 GHz, and above -10 dB for the whole bandwidth of the  bandgap. (b) Experimental setup used for measuring the near fields. Simulated (c) and measured (d) real electric field $\sim$1 mm above the structure.}
	\label{fig4}
\end{figure}

To verify the design, a sample of the structure was fabricated and measured, using the same parameters as given previously (with the addition of the aforementioned exponential taper to microstrip line, shown in Fig. \ref{fig4}(b)). Fig. \ref{fig4}(a) shows the comparison between simulated and measured scattering parameters, where the agreement is strong for the majority of the bandgap. Note that the simulation (solid lines) is for the antipodal fed structure (including all forms of loss), while the experimentally measured data shows the full structure with taper, minus the contribution from the taper. When the taper is included, an additional ~2 dB of loss is added to the shown values. Near 16.2 GHz, the transmission is above -1 dB, in close agreement with the simulated values.

For visual confirmation, the near field electric fields were measured via a 2D scanner with the probe tip placed $\sim$1 mm above the surface. Fig. \ref{fig4}(c) and (d) show the comparison between the simulated and measured real electric fields. Even with the tight vertical confinement it is clear that the desired nontrivial edge mode is strongly excited.

As discussed previously, a chief interest in PTI designs is their immunity to backscattering when presented with a wide class of defects and deformations to their propagation channel. Examples of these are sharp bends, removal of unit cells, and deformations to structure. However, it is often not clear to what extent these robust features present a practical benefit over more common transmission lines and waveguiding technology. With the utility of the efficient transition presented here, it is possible to quantify the performance vs backscatter presented in many implementations. 

The most common test for robustness in a PTI is the sharp bend, where the waveguide follows along the 60 or 120 degree bends present in the lattice \cite{he_silicon--insulator_2019}. In ordinary transmission lines such angles present substantial backscatter, and must be dealt with individually via miter or other bend engineering. Such methods frequently work in microwave devices by reducing the capacitance of the bend, and there are various empirically derived rules of thumb to compute the needed parameters \cite{douville_experimental_1978}. However, PTIs present an interesting design possibility, as such bends are automatically "immune" from scattering at any angle that fits within its lattice, as such modes cannot smoothly convert into those moving in the other direction. 
\begin{figure}
	\centering
	\includegraphics[width = \linewidth]{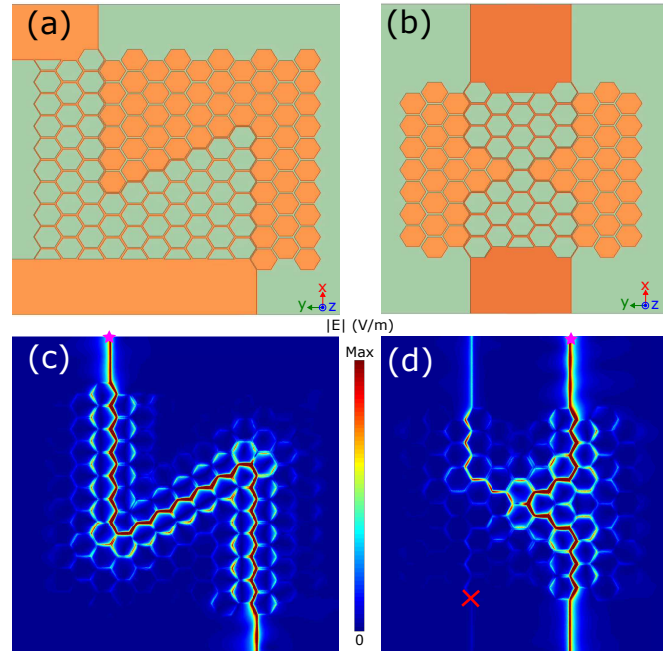}
	\caption{(a) Simulated sharp bending path, with two 120 degree turns. (b) Simulated magic T structure, with four input/output ports. (c) Electric field profile for (a), showing negligible losses. The excitation is marked with a pink star. (d) Electric field profile for (b), with port 1 (denoted by the pink star) excited. The downward side has high transmission, while there is practically zero in the "forward" channel. The asymmetry at the gap causes the upward side to have lower transmission, but still well above the "forward" channel.}
	\label{fig5}
\end{figure}

Fig. \ref{fig5}(a) shows a model of the duality PTI with the presented coupler to a classical ASL, with its field profile shown in Fig. \ref{fig5}(c). There is no major scattering visible, but the presented transition enables further analysis of the losses. Using the values in Table I we can calculate the losses incurred by the bends separately. For the model shown we calculate a loss of 0.036 db per bend (0.073 dB for both). Since the bending losses are close to the simulation accuracy, they can be considered negligible compared to other sources (i.e., dielectric, radiation, and transition). Being an inherent property of  topological protection, such a device can be scaled to much higher frequencies (e.g., into the the mmWave band of 5G telecommunications systems) without changing the device design. This is in contrast to classical waveguides, where increasing frequency often degrades the performance of sharp bend compensation methods, thus requiring re-engineering.

The other major demonstration of PTI devices is the so-called "magic T," which is a four-way junction of non-trivial waveguides \cite{makwana_designing_2018}. In a classical rectangular waveguiding system, such a junction results in equal energy splitting between the two side channels and no transmission through the forward channel. It is significantly harder to form such a junction in planar transmission lines without careful engineering \cite{u-yen_broadband_2008}. Due to the spin-momentum locking feature of PTIs, however, such behavior is intrinsic to such structures, even in the case of large mismatches of the propagation direction \cite{cheng_robust_2016}. 

Such a setup is shown in Fig. \ref{fig5}(b), with the corresponding electric field profile for port 1 activated shown in Fig. \ref{fig5}(d). Note that the hexagonal unit cells cause there to be a jump in the left-right symmetry of the waveguiding channel at the interface, which causes a drop in signal across the gap. Nevertheless, there is more than a 20 dB difference in transmission between the forward channel and the oppositely-oriented side channel. The transmission into the forward channel is close to -30 dB, despite being well matched to the direction of propagation compared to the side channel, which is -11 dB. It is possible to manually modify the interface region (e.g., via collapsing the center cell) which allows for greater similarity of the energy split to each side channel, but comes at the expense of increased signal in the forward (forbidden) port. This is due to spin flip processes that allow for energy coupling between the two pseudo-spin modes. See the Supplementary Material for further details.

The coupler presented here compares favorably with similar slow-wave devices. The SPP designs of \cite{ma_broadband_2014}, \cite{kianinejad_design_2015}, and \cite{yan_broadband_2020} report $S_{21}$ values of -0.6 dB, -1.5 dB, and -2.0 dB, respectively, for most of their bandwidth, which are all close to or below the -0.866 dB reported here. Moreover, the topological protection afforded to the design allows for  negligible losses under sharp bends like those of Fig \ref{fig5}(c), which would otherwise cause substantial scattering for SPP designs. Likewise, the propagation losses reported for the PTI structure itself are comparable to those of standard microwave devices (e.g., a 50 $\Omega$ microstrip would have $\sim$0.032 dB/cm vs the $\sim$0.031 dB/cm for the PTI design with both at 16.2 GHz and equal material settings \cite{hammerstad_accurate_1980}), all without the need for gradual bends or engineered corners \cite{douville_experimental_1978}.

This Letter presents a compact transition for efficiently coupling energy from a traditional planar transmission line to a topologically protected line wave. An antipodal slot line was used as the initial transmission line, which displays a field profile closely related to the PTI structure. Since the PTI's E-field distribution varies substantially along a unit cell, the exact coupling point is chosen for the greatest match to the antipodal line. Additional alterations to the abrupt transition to account for lattice mismatch and edge coupling effects improve the design further. Use of such a design permits quantitative analyses of many popularly reported structures that rely upon topological phases in waveguiding applications. The resulting transition allows for less than 2.1\% /transition of added loss, bringing topologically robust structures closer to integration in practical applications. 

\section*{Data Availability}
The data that support the findings of this study are available from the corresponding authors upon reasonable request.
\section*{Supplementary Material}
See supplementary material for details on the effects of finite bulk sizes, and for extended data on the results shown in Fig. 5. 
\begin{acknowledgments}
This work was supported by AFOSR grant FA9550-16-1-0093.
\end{acknowledgments}

\bibliography{cites}

\end{document}